\documentclass[aps,pra,twocolumn,groupedaddress]{revtex4-1}

\usepackage{epsfig}
\usepackage{graphicx}
\usepackage{dcolumn}
\usepackage{bm}
\usepackage{amsthm}
\usepackage{amsfonts}
\usepackage{amscd}
\usepackage{amsmath}    
\usepackage{enumerate}

\newcommand{\ket}[1]{\mbox{$ | #1 \rangle $}}
\newcommand{\bra}[1]{\mbox{$ \langle #1 | $}}

\begin{document}


\title{A passive transmitter for quantum key distribution with coherent light}



\author{Marcos Curty$^1$, Marc Jofre$^2$, Valerio Pruneri$^{2,3}$, and Morgan W. Mitchell$^2$}
\affiliation{$^1$ Escuela de Ingenier\'ia de Telecomunicaci\'on, Department of Signal Theory and Communications, University of Vigo, 
Campus Universitario, 36310 Vigo, Pontevedra, Spain \\
$^2$ ICFO-Institut de Ci\`encies Fot\`oniques, Mediterranean Technology 
Park, 08860 Castelldefels, Barcelona, Spain \\
$^3$ ICREA-Instituci\'o Catalana de Recerca i Estudis Avanc\c{}ats, 08010 Barcelona, Spain
}


\date{\today}

\begin{abstract}
Signal state preparation in quantum key distribution schemes can be realized using either an active or a passive source. 
Passive sources might be valuable in some scenarios; for instance, in those experimental setups operating at 
high transmission rates, since no externally driven element is required. Typical passive transmitters involve 
parametric down-conversion. More recently, it has been shown that 
phase-randomized coherent pulses also allow 
passive generation of decoy states and 
Bennett-Brassard 1984 (BB84) polarization signals, though 
the combination of both setups in a single passive source is cumbersome. In this paper, 
we present a complete passive transmitter that prepares decoy-state BB84 signals using coherent light. 
Our method employs sum-frequency generation together with linear optical components and classical photodetectors. In the 
asymptotic limit of an infinite long experiment, the resulting 
secret key rate (per pulse) 
is comparable to the one delivered by an active decoy-state BB84 setup with 
an infinite number of decoy settings.
\end{abstract}


\maketitle

\section{Introduction}

Quantum key distribution (QKD) is already a mature technology that can provide
cryptographic systems with an unprecedented level of security \cite{qkd}. 
It aims at the distribution of a secret key between 
two distant parties (typically called Alice and Bob)
despite the technological power 
of an eavesdropper (Eve) who interferes with the signals. This secret key 
is the essential ingredient of the one-time-pad or Vernam 
cipher \cite{Vernam}, the only known encryption method that can offer
information-theoretic secure communications.

Most practical long-distance implementations of QKD are based on the so-called BB84 protocol,
introduced by Bennett and Brassard in 1984 \cite{bb84}, in combination with the 
decoy-state method \cite{decoy,decoy2,decoy2b,model2,decoy_e}.
In a typical quantum 
optical implementation of this scheme, Alice sends to Bob phase-randomized weak coherent pulses (WCPs) with
different mean photon numbers that are selected, 
independently and randomly, for each signal. These states can be generated using 
a standard semiconductor laser together with a variable optical attenuator that is controlled by a 
random number generator (RNG) \cite{random}.
Each light pulse may be prepared in a different polarization state, which is selected, 
again independently and randomly for each signal, between 
two mutually unbiased bases, 
{\it e.g.}, either a linear (H [horizontal] or V [vertical]) or 
a circular (L [left] or R [right]) polarizations basis \cite{note}. For that, two main experimental 
configurations are typically used. 
In the first one, Alice employs 
four laser diodes, one for each possible BB84 signal \cite{4lasers}. These lasers are controlled 
by a RNG that decides each given time which one of the four diodes is triggered. 
The second configuration utilizes only one laser diode 
in combination with a polarization modulator \cite{modulator}.
This modulator rotates the state of polarization of the signals 
depending on the output of a 
RNG. On the receiving side, Bob measures each incoming signal by choosing at random 
between two polarization analyzers, one for each
 possible basis. Once the quantum communication phase of the protocol is completed, Alice and Bob use an authenticated public 
channel to process their data and obtain a secure secret key. 
This last procedure, called key distillation, involves, generally, 
local randomization, error correction to reconcile Alice's and Bob's data, and privacy ampliÞcation to 
decouple their data from Eve \cite{post}. A full proof of the security for the decoy-state BB84 QKD protocol with WCPs 
has been given in Refs.~\cite{decoy2,decoy2b,model2}.

Alternatively to the active signal state preparation methods described above, 
Alice may as well employ a passive transmitter to generate decoy-state 
BB84 signal states. This last solution might be desirable in some scenarios; 
for instance, in those experimental setups operating at high transmission rates, 
since no RNGs are required in a passive device \cite{rarity,passive1,passive2,passive3,passive4,passiveEB}. 
Passive schemes could also be more robust against 
side-channel attacks hidden in the imperfections of the optical components
than active 
sources. The working principle of a 
passive transmitter is rather simple. 
For example, Alice can use various light sources 
to produce different signal states that are sent through an optics network. Depending on the 
detection pattern observed in some properly located photodetectors, she can infer which signal states are 
actually generated. Known passive schemes rely typically on the use of a parametric down-conversion (PDC) source, 
where Alice and Bob passively and randomly choose which bases to measure each incoming pulse by means of a 
beamsplitter (BS) \cite{passiveEB}. 
Also, Alice can exploit the photon number correlations that exist between the two output modes of a PDC source to 
passively generate 
decoy states \cite{passive1}. 
More recently, it has been shown that phase-randomized coherent pulses are also suitable 
for passive preparation of decoy states \cite{passive2,passive3} and BB84 polarization signals \cite{curty}, though 
the combination of both setups in a single passive source is cumbersome. Intuitively speaking, Refs.~\cite{passive2,passive3,curty} 
take advantage of the random phase of the different incoming pulses to 
passively generate states with either distinct photon number statistics but with the same polarization \cite{passive2,passive3}, 
or with different polarizations but equal 
intensities \cite{curty}.

In this article, 
we present a complete passive transmitter for QKD that can prepare decoy-state BB84 signal states using coherent light. 
Our method employs sum-frequency generation (SFG) \cite{sfg1,kumar} together with linear optical components and classical photodetectors. 
SFG has already exhibited its 
usefulness in quantum information \cite{sfg2} and device-independent QKD \cite{sfg3} at the single-photon level. Here we use it in the 
conventional non-linear optics paradigm with strong coherent light. 
This fact might render our proposal particularly valuable from an experimental point of view. 
In the asymptotic limit of an infinite long experiment, it turns out that the secret 
key rate (per pulse) provided by such passive scheme is similar to the 
one delivered by an active decoy-state BB84 setup with infinite decoy 
settings. 

The paper is organized as follows. In Sec.~\ref{pol} we introduce a passive transmitter that 
generates 
decoy-state 
BB84 polarization signal states 
using coherent light. Then, in Sec.~\ref{sec_lower}
we evaluate its performance and we obtain a lower bound on the resulting 
secret key rate. In Sec.~\ref{phase} we consider the case where Alice and Bob use phase-encoding, which is more suitable to 
employ in combination with optical fibers than polarization encoding. 
Finally, Sec.~\ref{conc} concludes the article with 
a summary. The paper includes as well some Appendixes 
with additional calculations.

\section{Passive decoy-state BB84 transmitter}\label{pol}

\begin{figure}
\begin{center}
\includegraphics[angle=0,scale=0.65]{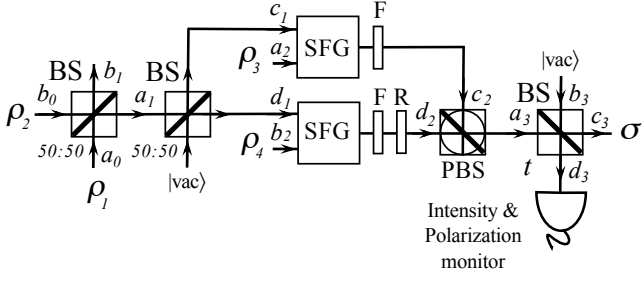}
\end{center}
\caption{Basic setup of a passive decoy-state BB84 QKD source with polarization encoding using 
phase-randomized strong coherent pulses. The mean photon number of the signal states 
$\rho_i$, with $i\in\{1,\ldots,4\}$, can be chosen very high; for instance,
$\approx{}10^8$ photons. BS denotes a beamsplitter, PBS represents a 
polarizing beamsplitter in the $\pm45^\circ$ linear polarization basis \cite{note_last}, 
F is an optical filter,
R denotes a polarization rotator changing $+45^\circ$ linear polarization to $-45^\circ$ linear polarization,
$\ket{\rm vac}$ represents the 
vacuum state,
and $t$ denotes the transmittance of a BS; it satisfies 
$t\ll{}1$.
\label{figure_general}}
\end{figure}

The basic setup is illustrated in Fig.~\ref{figure_general}.
Let us start considering, for simplicity, the interference of 
two pure coherent states of frequency $w_1$, 
both prepared in $+45^\circ$ linear
polarization and
with arbitrary phase relationship, $\ket{\sqrt{2\mu}e^{i\theta_1}}_{a_0,+45^\circ}$ and 
$\ket{\sqrt{2\mu}e^{i\theta_2}}_{b_0,+45^\circ}$, at a $50:50$ BS. The
output states in modes $a_1$ and $b_1$ are given by
\begin{equation}
\ket{\sqrt{\mu}(e^{i\theta_1}+e^{i\theta_2})}_{a_1,+45^\circ}\otimes\ket{\sqrt{\mu}(e^{i\theta_1}-e^{i\theta_2})}_{b_1,+45^\circ}.
\end{equation}
Then, we have that the output states in modes $c_1$ and $d_1$ have the form
\begin{equation}
\ket{\sqrt{\frac{\mu}{2}}(e^{i\theta_1}+e^{i\theta_2})}_{c_1,+45^\circ}\otimes\ket{\sqrt{\frac{\mu}{2}}(e^{i\theta_1}+e^{i\theta_2})}_{d_1,+45^\circ}.
\end{equation}
If these two states
are combined with two coherent states of frequency $w_2$,
 $\ket{\sqrt{\mu}e^{i\theta_3}}_{a_2,+45^\circ}$ and 
$\ket{\sqrt{\mu}e^{i\theta_4}}_{b_2,+45^\circ}$, 
in a nonlinear medium using the SFG process,
the resulting output states at frequency $w_3=w_1+w_2$, 
after the polarization rotation $R$,
can be written as
(see Appendix~\ref{ap1})
\begin{equation}\label{sfg_eq}
\ket{\frac{-\sqrt{\mu}e^{i\theta_3}(e^{i\theta_1}+e^{i\theta_2})}{\sqrt{2}}}_{c_2,+45^\circ}
\otimes
\ket{\frac{-\sqrt{\mu}e^{i\theta_4}(e^{i\theta_1}+e^{i\theta_2})}{\sqrt{2}}}_{d_2,-45^\circ}.
\end{equation}
These two beams are now re-combined at a PBS in the $\pm45^\circ$ linear polarization basis \cite{note_last}. We obtain that the output state
in mode $a_3$ (see Fig.~\ref{figure_general}) is a coherent state of the form
\begin{equation}\label{tor25}
\ket{\sqrt{\zeta(\theta)}e^{i\phi}}_{\psi,a_3}=e^{-\zeta(\theta)/2}\sum_{n=0}^\infty \frac{\big(\sqrt{\zeta(\theta)}e^{i\phi}\big)^n}{\sqrt{n!}}
\ket{n_\psi},
\end{equation}
where $\zeta(\theta)=2\mu(1+\cos{\theta})$, $\theta=\theta_2-\theta_1$, $\phi=\pi+\theta_1+\theta_3+\arg{(1+e^{i\theta})}$, 
and the Fock states $\ket{n_\psi}$ are given by
\begin{equation}
\ket{n_\psi}=\frac{[\frac{1}{\sqrt{2}}(a_{+45^\circ}^\dagger+e^{i\psi}a_{-45^\circ}^\dagger)]^n}
{\sqrt{n!}}\ket{\rm vac},
\end{equation}
with $\ket{\rm vac}$ denoting the vacuum state and $\psi=\theta_4-\theta_3$. Finally, 
Alice sends 
the quantum state given by Eq.~(\ref{tor25})
through a BS of transmittance $t\ll{}1$. Then, the output 
states in modes $c_3$ and $d_3$ are given by
\begin{equation}\label{eq_bf}
\ket{\sqrt{t\zeta(\theta)}e^{i\phi}}_{\psi,c_3}
\otimes
\ket{\sqrt{(1-t)\zeta(\theta)}e^{i\phi}}_{\psi,d_3}.
\end{equation}

The analysis of the case where the global phase of each input signal $\rho_i$, with $i\in\{1,\ldots,4\}$, 
is randomized and inaccessible to the eavesdropper is now straightforward.  
It can be solved by just integrating the 
signals $\ket{\sqrt{t\zeta(\theta)}e^{i\phi}}_{\psi,c_3}$ and $\ket{\sqrt{(1-t)\zeta(\theta)}e^{i\phi}}_{\psi,d_3}$ 
given by Eq.~(\ref{eq_bf})
over all angles $\theta,\phi$, and $\psi$. In particular, we have that 
the output state $\sigma$ in this scenario (see Fig.~\ref{figure_general}) can be written as
\begin{eqnarray}
\sigma&=&\frac{1}{(2\pi)^3}\iiint_{\phi,\theta,\psi} 
\ket{\sqrt{t\zeta(\theta)}e^{i\phi}}_{\psi,c_3}\bra{\sqrt{t\zeta(\theta)}e^{i\phi}}
\ {\rm d}\phi{\rm d}\theta{\rm d}\psi \nonumber \\
&=&\frac{1}{(2\pi)^2}\int_\theta e^{-\gamma(\theta)}\sum_{n=0}^\infty \frac{\gamma(\theta)^n}
{n!}\int_\psi \ket{n_\psi}\bra{n_\psi} \ {\rm d}\theta{\rm d}\psi,
\end{eqnarray}
where the intensity $\gamma(\theta)$ is given by $\gamma(\theta)=t\zeta(\theta)$.

The weak intensity signal $\sigma$ in mode $c_3$ is suitable for QKD and Alice sends it to Bob through the 
quantum channel. Also,  
she uses the strong intensity signal available in mode $d_3$ to measure both 
its intensity and polarization. This last measurement can be realized, 
for example, by means of a passive BB84 detection scheme where the 
basis choice is performed by a $50:50$ BS, and on each end there is a PBS
and two classical photodetectors. From the different intensities observed 
in each of these four photodetectors, Alice can determine both the value 
of the angle $\psi$ and the total intensity of the signal. Note that, by
assumption, we have that the intensity of the input states $\rho_i$ is very high. 

\begin{figure}
\begin{center}
\includegraphics[angle=0,scale=0.57]{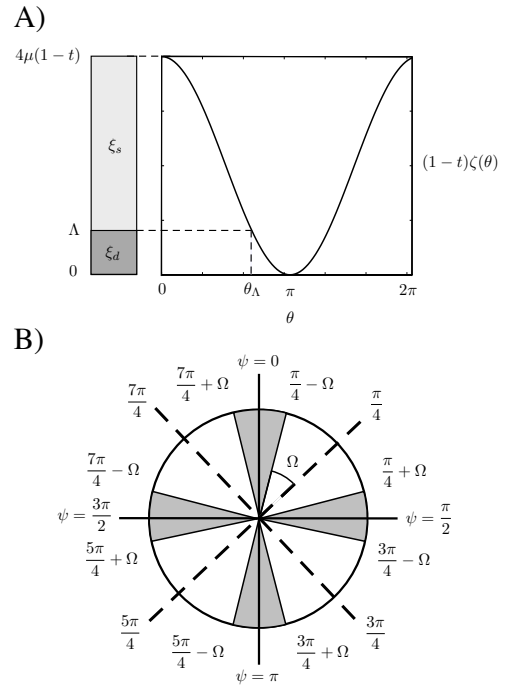}
\end{center}
\caption{(Case A) Graphical representation of the intensity $(1-t)\zeta(\theta)$ in mode $d_3$
(see Fig.~\ref{figure_general}) versus the angle $\theta$. $\Lambda$ represents the 
threshold value of the classical intensity measurement, $\theta_\Lambda$ is its associated 
threshold angle, and $\xi_d$ and $\xi_s$ denote the resulting intensity intervals.
(Case B) 
Graphical representation of the valid regions for the 
angle $\psi$. These regions are marked in gray. They depend on 
an acceptance parameter $\Omega\in[0,\pi/4]$.
\label{figure_intervals}}
\end{figure}

For simplicity, let us assume for the moment that 
the polarization measurement is perfect, {\it i.e.}, for each incoming signal it provides Alice 
with a precise value for the measured angle $\psi$,
while the intensity measurement only tells her whether the measured intensity
is below or above a certain threshold value $\Lambda$ that satisfies 
$0<\Lambda<4\mu(1-t)$. That is, $\Lambda$ is 
between the minimal and maximal possible values of the intensity $(1-t)\zeta(\theta)$ of the optical pulses
in mode $d_3$. The first intensity interval, $\xi_d=[0,\Lambda]$, can be associated, for instance, to the generation of a decoy state in 
output mode $c_3$ (that 
we shall denote as $\sigma_d$), while the second intensity interval, $\xi_s=[\Lambda,4\mu(1-t)]$, corresponds to the case of preparing a signal state ($\sigma_s$). 
Note, however, that the analysis presented in this section 
can be straightforwardly adapted to cover as well the case of several intensity 
intervals $\xi_i$ ({\it i.e.}, the generation of several decoy states). Figure~\ref{figure_intervals} (case A) shows a graphical
representation of the intensity $(1-t)\zeta(\theta)$ in mode $d_3$ versus the angle $\theta$, together with the 
threshold value $\Lambda$ and the intensity intervals $\xi_d$ and $\xi_s$. 
The threshold angle $\theta_\Lambda$
that satisfies $(1-t)\zeta(\theta_\Lambda)=\Lambda$ is given by
\begin{equation}
\theta_\Lambda=\arccos{\Bigg(\frac{\Lambda}{2\mu(1-t)}-1\Bigg)}.
\end{equation}

In this simplified scenario, the 
conditional quantum states that are sent to Bob can be written as
\begin{equation}\label{sig}
\sigma_{i,\psi}=\sum_{n=0}^\infty p_n^i \ket{n_\psi}\bra{n_\psi},
\end{equation}
where $i=\{s,d\}$, and
the probabilities $p_n^{i}$ are given by
\begin{eqnarray}\label{prob}
p_n^{s}&=&\frac{1}{\theta_\Lambda}\int_{0}^{\theta_\Lambda} 
e^{-\gamma(\theta)}\frac{\gamma(\theta)^n}{n!} \  {\rm d}\theta, \nonumber \\
p_n^{d}&=&\frac{1}{\pi-\theta_\Lambda}\int_{\theta_\Lambda}^\pi
e^{-\gamma(\theta)}\frac{\gamma(\theta)^n}{n!} \  {\rm d}\theta.
\end{eqnarray}

In practice, however, it is not necessary that Alice determines the value of $\psi$ accurately 
and 
restricts herself to 
only those events where she actually prepares a perfect BB84 polarization state
({\it i.e.}, when the angle $\psi$ satisfies $\psi\in\{0,\pi/2,\pi,3\pi/2\}$) \cite{curty}.
Note that the probability 
associated with these ideal events tends to zero. Instead, it is sufficient if the
polarization measurement tells her the value of $\psi$ within a certain interval 
around the desired ideal values. This situation is illustrated 
in Fig.~\ref{figure_intervals} (case B), where Alice selects some valid regions (marked with gray color in the figure) for the 
angle $\psi$ \cite{curty}. These regions depend on an acceptance parameter $\Omega\in[0,\pi/4]$ that we optimize. 
In particular, whenever the value of $\psi$ lies within any of the valid regions, 
Alice considers the pulse emitted by the source as a valid signal. 
Otherwise, the pulse is discarded afterwards during the post-processing phase of the protocol, and it 
does not contribute to the key rate. The probability that a pulse is accepted, $p_{\rm acc}$,
is given by 
\begin{equation}
p_{\rm acc}=1-\frac{4\Omega}{\pi}.
\end{equation}

There is a trade-off on the acceptance parameter $\Omega$. A 
high acceptance probability $p_{\rm acc}$ favors $\Omega\approx{}0$, but this action 
also results in an increase of the quantum bit error rate (QBER) of the protocol. A 
low QBER favors $\Omega\approx{}\pi/4$, but then $p_{\rm acc}\approx{}0$. 
Note that in the limit where 
$\Omega$ tends to $\pi/4$ we recover the standard decoy-state BB84 protocol.

\section{Lower bound on the secret key rate}\label{sec_lower}

We shall consider that Alice and Bob treat 
decoy and signal states separately, and they distill secret key 
from both of them. For that, we use the security analysis presented in 
Ref.~\cite{decoy2}, which combines the results provided by 
Gottesman-Lo-L\"utkenhaus-Preskill (GLLP) in Ref.~\cite{sec_bb84b}  
(see also Ref.~\cite{hkl}) with the decoy-state method \cite{notetor}. 
The secret key rate formula can be written as
\begin{equation}\label{key}
R\geq{}\sum_i p_i\ {\rm max}\{R^i,0\},
\end{equation}
with $i=\{s,d\}$. Here $p_i$ denotes the probability to generate a state associated to the 
intensity interval $\xi_i$ ({\it i.e.}, $p_s=\theta_\Lambda/\pi$ and
$p_d=1-p_s$), and
\begin{eqnarray}\label{rate}
R^i&\geq&{}qp_{\rm acc}\big\{-Q^if(E^i)H(E^i)+p_1^iY_1[1-H(e_1)]\nonumber \\
&+&p_0^iY_0\big\}.
\end{eqnarray}
The parameter $q$ is the efficiency of the protocol ($q=1/2$ for the standard BB84 scheme, and 
$q\approx{}1$ for its efficient version \cite{eff}); $Q^i$ denotes the gain, {\it i.e.}, the 
probability that Bob obtains a click in his measurement apparatus when Alice 
sends him a signal $\sigma_i$; 
$f(E^i)$ represents the efficiency 
of the error correction protocol as a function of the error rate $E^i$, typically $f(E^i)\geq{}1$
with Shannon limit $f(E^i)=1$ \cite{gb}; $Y_n$ is the yield of an $n$-photon signal, 
{\it i.e.}, the conditional probability of a detection event on Bob's side 
given that Alice transmits an $n$-photon state; 
$e_n$ denotes the error rate of an $n$-photon signal; 
and $H(x)=-x\log_2{(x)}-(1-x)\log_2{(1-x)}$ represents the binary Shannon entropy function. 

For simulation purposes, we shall consider a simple channel model in the absence of
eavesdropping \cite{decoy2,model2}; it just consists of a BS whose transmittance
depends on the transmission distance and the loss coefficient of the quantum channel.
That is, for simplicity, we neglect any misalignment effect in the channel.
Furthermore, we assume 
that Bob employs an
active BB84 detection setup. This model allows us to calculate
the observed experimental parameters $Q^i$ and $E^i$. These quantities are given 
in Appendix~\ref{gqber}. Our results, however, can also be straightforwardly 
applied to any other quantum channel or detection setup, as they depend only 
on the observed gain and QBER. 

To evaluate the secret key rate formula given by Eq.~(\ref{rate}) we need to estimate the 
yields $Y_0$ and $Y_1$, together with the single-photon error rate $e_1$, 
by solving the following set of linear equations:
\begin{equation}
Q^i=\sum_{n=0}^\infty p_n^i Y_n, \quad {\rm and} \quad
Q^iE^i=\sum_{n=0}^\infty p_n^i Y_ne_n.
\end{equation} 
For that, we shall use the procedure proposed in 
Refs.~\cite{model2,passive3}. 
Moreover, we will assume a random background 
({\it i.e.}, $e_0=1/2$).
This method requires that the probabilities
$p_n^i$ given by Eq.~(\ref{prob}) satisfy certain conditions that we confirm numerically. 
The results
are included in Appendix~\ref{estimation}.
It is important to emphasize, however, that the estimation 
technique presented in 
Refs.~\cite{model2,passive3} 
only constitutes a possible example of a finite setting estimation 
procedure. In principle, many other 
estimation methods are also available for this purpose, such as linear programming tools \cite{linear},
which might result in sharper, or for the purpose of QKD, better bounds on the 
considered probabilities. 

\begin{figure}
\begin{center}
\includegraphics[angle=0,scale=0.35]{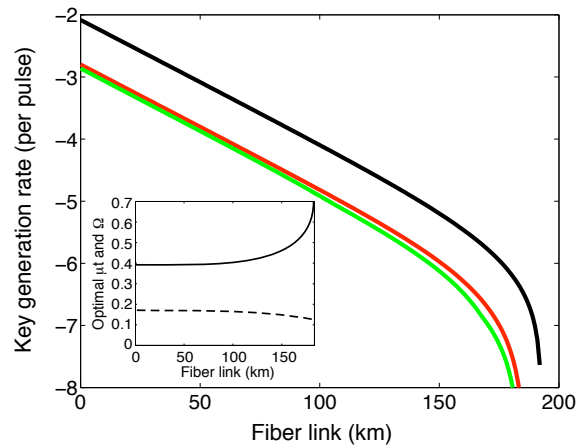} 
\end{center}
\caption{Lower bound on the secret key rate $R$ given by 
Eq.~(\ref{key}) in logarithmic scale for the passive transmitter 
with two intensity settings illustrated 
in Fig.~\ref{figure_general} (green line). 
For simulation 
purposes, we consider the following experimental parameters:
the dark count rate of Bob's detectors is
 $\epsilon_{\rm B}=3.2\times10^{-7}$, the overall transmittance of Bob's detection
 apparatus is 
 $\eta_{\rm B}=0.045$, the loss coefficient of the channel is 
$\alpha=0.2$ dB/km, $q=1/2$, and the efficiency of the error correction protocol is
$f(E^i)=1.22$. 
We further assume the channel 
model described in Refs.~\cite{decoy2,model2},
where we neglect any misalignment 
effect. Otherwise, the actual secure distance will be smaller.
The inset figure shows the value for the optimized parameters $\mu{}t$ (dashed line) and
$\Omega$ (solid line) in the passive setup. The optimal value for the threshold parameter $\Lambda$ turns out to be 
constant with the distance and equal to
$2\mu(1-t)$, {\it i.e.}, the threshold angle $\theta_\Lambda$ satisfies $\theta_\Lambda=\pi/2$.
The black line represents a lower bound on $R$ for an active asymptotic decoy-state 
BB84 system with infinite decoy settings \cite{decoy2}, while the red line shows the 
case of a passive transmitter with infinite intensity intervals $\xi_i$ (see Appendix~\ref{asymptotic}). 
\label{figrate}}
\end{figure}

The resulting lower bound on the secret key rate with two intensity settings is illustrated
in Fig.~\ref{figrate} (green line). 
In our simulation we employ the following experimental parameters: 
the dark count rate of Bob's detectors is
$\epsilon_{\rm B}=3.2\times10^{-7}$, 
the overall transmittance of Bob's detection
 apparatus is 
$\eta_{\rm B}=0.045$, 
and the loss coefficient of the channel is 
$\alpha=0.2$ dB/km.
We further assume that 
 $q=1/2$, and $f(E^i)=1.22$. With this configuration, it turns out that 
 the optimal value of the parameter $\mu{}t$ decreases with increasing distance, 
 while the optimal
 value of the parameter $\Omega$ increases with the distance. A similar behavior 
was also observed in the passive BB84 transmitter (without decoy states) proposed in Ref.~\cite{curty}. 
 In particular, $\mu{}t$ diminishes from $\approx{}0.175$ to $\approx{}0.125$, while
 $\Omega$ augments from $\approx{}0.393$ to $\approx{}0.7$. At long distances 
 the gain of the protocol is very low and, therefore, it is important to keep both the 
 multi-photon probability of the source (related with the parameter $\mu{}t$) and the intrinsic error rate of the signals 
 sent by Alice (related with the 
 parameter $\Omega$) also low. Figure~\ref{figrate} includes as well an inset 
plot with the optimized parameters $\mu{}t$ (dashed line) and $\Omega$ (solid line). The optimal 
value for the parameter $\Lambda$ turns out to be constant with the distance; it is given by $\Lambda=2\mu(1-t)$, 
{\it i.e.}, the threshold angle $\theta_\Lambda$ is equal to $\pi/2$. This figure also shows 
a lower bound on the secret key rate for the cases of an 
active decoy-state BB84 system with infinite decoy settings (black line) \cite{decoy2}, and a 
passive transmitter with infinite intensity intervals $\xi_i$ (red line). The cutoff points where the 
secret key rate drops down to zero are $\approx{}181$ km (passive setup 
with two intensity settings), $\approx{}183$ km (passive setup 
with infinite intensity settings), and $\approx{}192$ km (active transmitter 
with infinite decoy settings). 
From the results 
shown in Fig.~\ref{figrate} we see that the performance of the passive transmitter 
presented in Sec.~\ref{pol}, with only two intensity settings, is similar 
to that of an active asymptotic setup, thus showing the practical interest of the passive scheme. 
The relatively small difference between the achievable secret key rates in both scenarios
is due to two main factors: (a) the intrinsic error rate of the signals accepted by Alice, which is zero 
only in the case of an active source, and (b)
the probability $p_{acc}$ to accept a pulse emitted by the source, which is 
$p_{acc}<1$ in the passive setup and $p_{acc}=1$ in the active scheme. For instance, 
we have that
for most 
distances $\Omega\approx{}0.393$, which implies $p_{acc}\approx{}0.5$. This fact reduces 
the key rate on logarithmic scale of the passive transmitter by a factor of $\log_{10}{p_{acc}}\approx{}0.3$. 
The additional factor of $\approx{}0.45$ that can be observed
in Fig.~\ref{figrate}
arises mainly from the intrinsic error rate of the signals.

\section{Phase encoding}\label{phase}

Similar ideas to the ones presented in Sec.~\ref{pol} can also be used in other implementations 
of the decoy-state BB84 protocol with a different signal encoding. For instance, in those QKD experiments based on phase encoding, which 
is more suitable to use with optical fibers than polarization encoding, which is particularly relevant in the context of 
free-space QKD \cite{qkd}. 

\begin{figure}
\begin{center}
\includegraphics[angle=0,scale=0.62]{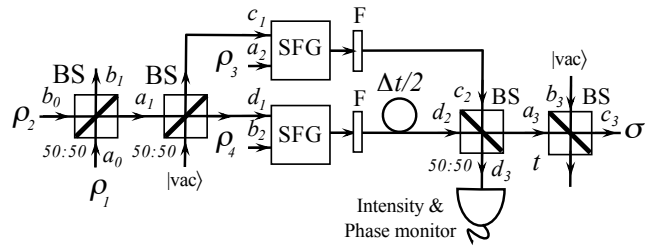}
\end{center}
\caption{Basic setup of a passive decoy-state BB84 QKD 
source with 
phase encoding. The delay introduced by one 
arm of the interferometer is equal to half the time difference $\Delta{}t$ between two consecutive pulses. 
\label{figure_general2}}
\end{figure}

The basic 
setup is illustrated in Fig.~\ref{figure_general2}. 
Again, for simplicity, let us consider first the case where the input signals $\rho_i$, with $i\in\{1,\ldots,4\}$, are pure coherent states with arbitrary 
phase relationship: $\ket{\sqrt{2\mu}e^{i\theta_1}}_{a_0,+45^\circ}$ and 
$\ket{\sqrt{2\mu}e^{i\theta_2}}_{b_0,+45^\circ}$ (of frequency $w_1$), and 
$\ket{\sqrt{\mu}e^{i\theta_3}}_{a_2,+45^\circ}$ and 
$\ket{\sqrt{\mu}e^{i\theta_4}}_{b_2,+45^\circ}$ (of frequency $w_2$). Let $\Delta{}t$
denote the time difference between two consecutive pulses generated by the sources.
Then, from Sec.~\ref{pol} we have that the signals in modes $c_2$ and $d_2$  
at time instances $t$ and $t+\Delta{}t/2$ can be written as
\begin{equation}
\ket{\sqrt{\frac{\zeta(\theta)}{2}}e^{i\phi}}_{c_2,+45^\circ}^t
\otimes
\ket{\sqrt{\frac{\zeta(\theta)}{2}}e^{i\phi'}}_{d_2,+45^\circ}^{t+\Delta{}t/2},
\end{equation}
where $\phi'=\phi+\theta_4-\theta_3$.
Similarly, we find that the quantum states in modes $c_3$ and $d_3$ are given by, respectively,
\begin{eqnarray}\label{imax}
\ket{\frac{\sqrt{\gamma(\theta)}}{2}e^{i\phi}}_{c_3,+45^\circ}^t
&\otimes&
\ket{\frac{\sqrt{\gamma(\theta)}}{2}e^{i\phi'}}_{c_3,+45^\circ}^{t+\Delta{}t/2}, \nonumber \\
\ket{\frac{\sqrt{\zeta(\theta)}}{2}e^{i\phi}}_{d_3,+45^\circ}^t
&\otimes&
\ket{\frac{\sqrt{\zeta(\theta)}}{2}e^{i\phi'}}_{d_3,+45^\circ}^{t+\Delta{}t/2}.
\end{eqnarray}

The case of phase-randomized strong coherent 
pulses is completely analogous to that of Sec.~\ref{pol} and we omit it 
here for simplicity; it results in an uniform 
distribution for the angles $\theta$, $\phi$, and $\phi'$ for both pairs of pulses
given by Eq.~(\ref{imax}).
The strong signals in mode $d_3$ are used to measure both their phases, 
relative to some local reference phase, and their intensities by means of an intensity 
and phase 
measurement, while Alice sends the weak signals in mode $c_3$ to Bob. Again, just like in 
the passive source with polarization encoding shown in Fig.~\ref{figure_general}, 
Alice can now select some valid 
regions for the measured phases and also distinguish between different intensity 
settings. Then, we have that the analysis and results presented 
in Sec.~\ref{sec_lower} also apply straightforwardly to this scenario.

\section{Conclusion}\label{conc}

In this paper, we have introduced a complete passive transmitter for QKD that can prepare decoy-state 
Bennett-Brassard 1984 signal states 
using coherent light. Our method employs sum-frequency generation together with linear optical components and classical photodetectors, 
and constitutes an alternative to those active sources that are typically used in current experimental realizations 
of QKD. In the asymptotic limit of an infinite long experiment, we have proven that such passive scheme
can provide a secret 
key rate (per pulse) similar to the 
one delivered by an active decoy-state BB84 setup with infinite decoy 
settings, thus showing the practical interest of the passive scheme. 

The main focus of this paper has been polarization-based 
realizations of the BB84 protocol, which are particularly 
relevant for free-space QKD. However, we have also shown
that similar ideas can as well be applied to other practical 
scenarios with different signal encodings, like, for instance, 
those QKD experiments based on phase encoding, which are 
more suitable for use in combination with optical fibers.

\section{Acknowledgments}
We thank F. Steinlechner for helpful discussions. 
M.C. thanks the University of Toronto for hospitality and support during his stay
in this institution. 
This work was supported by Xunta de Galicia, Spain (grant 
No. INCITE08PXIB322257PR), by 
ERDF funds under the project Consolidation of
Research Units 2008/075, 
and by the Ministerio de Educaci\'on y Ciencia, Spain (grants 
TEC2010-14832, FIS2007-60179, FIS2008-01051 and Consolider Ingenio CSD2006-00019).

\appendix

\section{Sum-frequency generation}\label{ap1}

For completeness, in this Appendix we include the calculations to derive Eq.~(\ref{sfg_eq}) in Sec.~\ref{pol}. Our starting point are the 
input states to one of the two SFG processes used in the passive transmitter illustrated in
Fig.~\ref{figure_general}: 
$\ket{\sqrt{\frac{\mu}{2}}(e^{i\theta_1}+e^{i\theta_2})}_{c_1,+45^\circ}$ and 
$\ket{\sqrt{\mu}e^{i\theta_3}}_{a_2,+45^\circ}$.
Such process is described by the Hamiltonian
$H=i\hbar\chi(c_1a_2c_2^\dagger-{\rm H.c.})$, where $c_2^\dagger$ represents the 
creation operator
for the light wave at frequency $w_3=w_1+w_2$ \cite{kumar}. 
The parameter $\chi$ is a coupling constant that is
proportional to the second-order susceptibility $\chi^{(2)}$ of the nonlinear material, and H.c. denotes 
a Hermitian conjugate. 
When the pump mode at frequency $w_2$ is kept strong and 
undepleted, then this mode can be typically treated classically as a complex number. 
With this assumption, we have that the effective Hamiltonian 
above can now be written as $H=i\hbar\chi(\sqrt{\mu}e^{i\theta_3}c_1c_2^\dagger-{\rm H.c.})$. Using the Heisenberg
equation of motion, it is straightforward to obtain the following coupled-mode equations:
\begin{equation}
\frac{{\rm d}c_1}{{\rm d}t}=-\chi\sqrt{\mu}e^{-i\theta_3}c_2, \quad \frac{{\rm d}c_2}{{\rm d}t}=\chi\sqrt{\mu}e^{i\theta_3}c_1,
\end{equation}
which can be solved in terms of initial values at $t=0$ to yield
\begin{eqnarray}
c_1(t)&=&c_1(0)\cos{(\sqrt{\mu}\chi{}t)}-e^{-i\theta_3}c_2(0)\sin{(\sqrt{\mu}\chi{}t)} \nonumber \\
c_2(t)&=&c_2(0)\cos{(\sqrt{\mu}\chi{}t)}+e^{i\theta_3}c_1(0)\sin{(\sqrt{\mu}\chi{}t)}.
\end{eqnarray}
At the point of complete conversion, $t_c=\pi/(2\sqrt{\mu}\chi)$, we obtain
\begin{equation}
c_1^\dagger(t_c)=-e^{i\theta_3}c_2^\dagger(0), \quad c_2^\dagger(t_c)=e^{-i\theta_3}c_1^\dagger(0).
\end{equation}
That is, at time $t_c$ we find that the resulting output state at frequency $w_3$ from the SFG process is given by
\begin{equation}
\ket{-\sqrt{\frac{\mu}{2}}e^{i\theta_3}(e^{i\theta_1}+e^{i\theta_2})}_{c_2,+45^\circ}.
\end{equation}

\section{Gain and QBER}\label{gqber}

In this Appendix, we obtain a mathematical expression for the 
observed gains $Q^i$ and error rates $E^i$, with $i\in\{s,d\}$, for the 
passive QKD transmitter with two intensity settings 
introduced in Sec.~\ref{pol}. For that, we employ the typical channel model 
in the absence of eavesdropping
\cite{decoy2,model2}; it just consists of a BS of transmittance 
$\eta_{\rm channel}=10^{-\frac{\alpha{}d}{10}}$, where
$\alpha$ denotes 
the loss coefficient of the channel measured in dB/km and $d$ is the 
transmission distance.
Moreover, for simplicity, we consider that Bob employs an active 
BB84 detection setup with two threshold detectors. 

The action of Bob's measurement device can be described
by two positive operator value measures (POVMs), one for 
each of the two BB84 polarization bases $\beta\in\{l,c\}$, 
with $l$ denoting a linear polarization basis and $c$ a circular 
polarization basis. 
Each POVM contains four elements: $G_{\rm vac}^\beta$, $G_{\rm 0}^\beta$,
$G_{\rm 1}^\beta$, and $G_{\rm dc}^\beta$. The first one 
corresponds to the case of no click in the detectors, the following two 
POVM operators give precisely one detection 
click, and the last one, 
$G_{\rm dc}^\beta$,
gives rise to both detectors being triggered.
These operators can be written as \cite{curty}
\begin{eqnarray}\label{gees}
G_{\rm vac}^\beta&=&[1-\epsilon_{\rm B}(2-\epsilon_{\rm B})]F_{\rm vac}^\beta, \nonumber \\
G_{\rm 0}^\beta&=&(1-\epsilon_{\rm B})\epsilon_{\rm B}F_{\rm vac}^\beta+(1-\epsilon_{\rm B})F_{\rm 0}^\beta, \nonumber \\
G_{\rm 1}^\beta&=&(1-\epsilon_{\rm B})\epsilon_{\rm B}F_{\rm vac}^\beta+(1-\epsilon_{\rm B})F_{\rm 1}^\beta, \nonumber \\
G_{\rm dc}^\beta&=&\openone-G_{\rm vac}^\beta-G_{\rm 0}^\beta-G_{\rm 1}^\beta.
\end{eqnarray}
Here we assume 
that the background rate is, to a good 
approximation, independent of the signal detection. Moreover, 
for easiness of notation, we consider only a background 
contribution coming from the dark count rate $\epsilon_{\rm B}$ of Bob's detectors and 
we neglect other background contributions like, for instance, 
stray light arising from timing pulses which are not completely 
filtered out in reception. 
The operators $F_{\rm vac}^\beta$, $F_{\rm 0}^\beta$,
$F_{\rm 1}^\beta$, and $F_{\rm dc}^\beta$
have the form 
\begin{eqnarray}\label{efes}
F_{\rm vac}^\beta&=&\sum_{n,m=0}^\infty (1-\eta_{\rm sys})^{n+m}\ket{n,m}_\beta\bra{n,m}, \nonumber \\
F_{\rm 0}^\beta&=&\sum_{n,m=0}^\infty [1-(1-\eta_{\rm sys})^{n}](1-\eta_{\rm sys})^m\ket{n,m}_\beta\bra{n,m}, \nonumber \\
F_{\rm 1}^\beta&=&\sum_{n,m=0}^\infty [1-(1-\eta_{\rm sys})^{m}](1-\eta_{\rm sys})^n\ket{n,m}_\beta\bra{n,m}, \nonumber \\
F_{\rm dc}^\beta&=&\sum_{n,m=0}^\infty [1-(1-\eta_{\rm sys})^{n}][1-(1-\eta_{\rm sys})^{m}] \nonumber \\
&\times&\ket{n,m}_\beta\bra{n,m}, 
\end{eqnarray}
with $\beta\in\{l,c\}$. The signals 
$\ket{n,m}_l$ ($\ket{n,m}_c$) represent the state which has $n$ photons in the horizontal (circular left) 
polarization mode and $m$ 
photons in the vertical (circular right) polarization mode. The parameter $\eta_{\rm sys}$ denotes the overall 
transmittance of the system. This quantity can be written as
$\eta_{\rm sys}=\eta_{\rm B}\eta_{\rm channel}$, where 
$\eta_{\rm B}$ is the overall transmittance of Bob's detection apparatus, {\it i.e.}, it includes
the transmittance of any optical component within Bob's measurement device together with the 
efficiency of his detectors.

In the scenario considered, 
it turns out that the gains $Q^i$ are independent of the 
actual polarization of the signals $\sigma_{i,\psi}$ given by Eq.~(\ref{sig}) and the basis $\beta$
used to measure them. We obtain
\begin{eqnarray}\label{gain} 
Q^i&=&1-{\rm Tr}\big(G_{\rm vac}^\beta\sigma_{i,\psi}\big)\nonumber \\
&=&1-
\frac{(1-\epsilon_{\rm B})^2}{p_i\pi}\int_{\theta_{\xi_i}} e^{-\eta_{\rm sys}\gamma(\theta)} \ {\rm d}\theta,
\end{eqnarray}
where $p_s=\theta_\Lambda/\pi$, $p_d=1-p_s$, $\theta_{\xi_s}=[0,\theta_\Lambda]$, 
and $\theta_{\xi_d}=[\theta_\Lambda,\pi]$. 

When $\theta_\Lambda=\pi/2$, which is the value that maximizes 
the secret key rate formula given by Eq.~(\ref{key}), we have that 
the gains $Q^i$ can be written as 
\begin{eqnarray}
Q^s&=&1-(1-\epsilon_{\rm B})^2A_-(\eta_{\rm sys}\zeta),
\nonumber \\
Q^d&=&1-(1-\epsilon_{\rm B})^2A_+(\eta_{\rm sys}\zeta),
\end{eqnarray}
where $\zeta=2\mu{}t$, and 
\begin{equation}
A_{\pm}(x)=e^{-x}\big[I_{0,x}\pm{}L_{0,x}\big].
\end{equation}
Here $I_{q,z}$ represents the modified Bessel function of the first kind, 
and $L_{q,z}$ denotes the modified Struve function. These functions are defined as \cite{Bessel}
\begin{eqnarray}
I_{q,z}&=&\frac{1}{2\pi{}i}\oint e^{(z/2)(t+1/t)} t^{-q-1} dt,  \\
L_{q,z}&=&\frac{z^q}{2^{q-1}\sqrt{\pi}\Gamma_{q+1/2}}\int_{0}^{\pi/2} \sinh{(z\cos{\theta}) \nonumber
\sin{\theta}^{2q}}d\theta.
\end{eqnarray}

The error rates $E^i$ depend on the value of the angle $\psi$. By symmetry, we
can restrict ourselves to evaluate the QBER in only one of the valid regions for $\psi$. Note that 
is the same in all of them. 
For instance, let us consider the case where 
$\psi\in[7\pi/4+\Omega, \pi/4-\Omega]$ (which corresponds to the horizontal 
polarization interval), and let $E_\psi^i$ denote the error rate of a signal 
$\sigma_{i,\psi}$
in that region. This quantity can be written as 
\begin{equation}\label{aimax}
E_\psi^i=\frac{1}{Q^i}{\rm Tr}\bigg[\bigg(G_1^l+\frac{1}{2}G_{\rm dc}^l\bigg)\sigma_{i,\psi}\bigg].
\end{equation}
Here we have considered the typical initial post-processing step in the BB84 
protocol, where double-click events are not discarded by Bob, but are randomly assigned to single-click
events. Equation~(\ref{aimax})
can be further simplified as
\begin{eqnarray}\label{uno}
E_\psi^i&=&\frac{1}{2Q^i}\big\{\epsilon_{\rm B}(\epsilon_{\rm B}-1)f_{0,\psi}^i+[2+\epsilon_{\rm B}(\epsilon_{\rm B}-3)]
f_{1,\psi}^i \nonumber \\
&+&(1-\epsilon_{\rm B})^2f_{{\rm dc},\psi}^i+\epsilon_{\rm B}(2-\epsilon_{\rm B})\big\},
\end{eqnarray}
where $f_{j,\psi}^i={\rm Tr}\big(F_j^l\sigma_{i,\psi}^i\big)$. After a short calculation, we obtain
\begin{eqnarray}
f_{0,\psi}^i&=&
\frac{1}{p_i\pi}\int_{\theta_{\xi_i}}
e^{-\eta_{\rm sys}\gamma(\theta)}\Big[-1+e^{\frac{1}{2}\eta_{\rm sys}\gamma(\theta)(1+\cos{\psi})}\Big] \ {\rm d}\theta, 
\nonumber \\
f_{1,\psi}^i&=&
\frac{1}{p_i\pi}\int_{\theta_{\xi_i}}
e^{-\eta_{\rm sys}\gamma(\theta)}\Big[-1+e^{\frac{1}{2}\eta_{\rm sys}\gamma(\theta)(1-\cos{\psi})}\Big] \ {\rm d}\theta, 
\nonumber \\
f_{{\rm dc},\psi}^i&=&1+
\frac{1}{p_i\pi}\int_{\theta_{\xi_i}}
e^{-\eta_{\rm sys}\gamma(\theta)}-
e^{-\frac{1}{2}\eta_{\rm sys}\gamma(\theta)(1-\cos{\psi})} \nonumber \\
&-&
e^{-\frac{1}{2}\eta_{\rm sys}\gamma(\theta)(1+\cos{\psi})}
\ {\rm d}\theta.
\end{eqnarray}
When $\theta_\Lambda=\pi/2$, these expressions can be simplified as
\begin{eqnarray}\label{dos}
f_{0,\psi}^s&=&-A_-(\eta_{\rm sys}\zeta)+A_-[\kappa_+(\psi)],
\\
f_{0,\psi}^d&=&-A_+(\eta_{\rm sys}\zeta)+A_+[\kappa_+(\psi)],
\nonumber \\
f_{1,\psi}^s&=&-A_-(\eta_{\rm sys}\zeta)+A_-[\kappa_-(\psi)],
\nonumber \\
f_{1,\psi}^d&=&-A_+(\eta_{\rm sys}\zeta)+A_+[\kappa_-(\psi)],
\nonumber \\
f_{{\rm dc},\psi}^s&=&1+A_-(\eta_{\rm sys}\zeta)-A_-[\epsilon_+(\psi)]
-A_-[\epsilon_-(\psi)], 
\nonumber \\
f_{{\rm dc},\psi}^d&=&1+A_+(\eta_{\rm sys}\zeta)-A_+[\epsilon_+(\psi)]
-A_+[\epsilon_-(\psi)],
\nonumber 
\end{eqnarray}
where the parameters $\kappa_\pm(\psi)$ and $\epsilon_{\pm}(\psi)$ have the form
\begin{eqnarray}\label{a2imax}
\kappa_\pm(\psi)&=&\eta_{\rm sys}\zeta[1-(1\pm\cos{\psi})/2], \nonumber \\
\epsilon_{\pm}(\psi)&=&\eta_{\rm sys}\zeta(1\pm\cos{\psi})/2.
\end{eqnarray}

The quantum bit error rates $E^i$ are then given by
\begin{equation}\label{tres}
E^i=\frac{2}{\pi-4\Omega}\int_{\frac{7\pi}{4}+\Omega}^{\frac{\pi}{4}-\Omega} E_{\psi}^i {\rm d}\psi.
\end{equation}
Combining Eqs.~(\ref{uno})-(\ref{dos}-\ref{tres}), we find that
\begin{widetext}
\begin{eqnarray}
E^s&=&\frac{1}{2Q^s}\Bigg\{
1-(1-\epsilon_{\rm B})^2A_-(\eta_{\rm sys}\zeta)+\frac{2}{\pi-4\Omega}\int_{\frac{7\pi}{4}+\Omega}^{\frac{\pi}{4}-\Omega}
\epsilon_{\rm B}(\epsilon_{\rm B}-1)A_-[\kappa_+(\psi)]+[2+\epsilon_{\rm B}(\epsilon_{\rm B}-3)]
A_-[\kappa_-(\psi)]
 \nonumber \\
&-&(1-\epsilon_{\rm B})^2\big[A_-[\epsilon_+(\psi)]
+A_-[\epsilon_-(\psi)]\big] {\rm d}\psi\Bigg\}, \nonumber \\ 
\end{eqnarray}
\end{widetext}
and
\begin{widetext}
\begin{eqnarray}
E^d&=&\frac{1}{2Q^d}\Bigg\{
1-(1-\epsilon_{\rm B})^2A_+(\eta_{\rm sys}\zeta)+\frac{2}{\pi-4\Omega}\int_{\frac{7\pi}{4}+\Omega}^{\frac{\pi}{4}-\Omega}
\epsilon_{\rm B}(\epsilon_{\rm B}-1)A_+[\kappa_+(\psi)]+[2+\epsilon_{\rm B}(\epsilon_{\rm B}-3)]
A_+[\kappa_-(\psi)]
 \nonumber \\
&-&
(1-\epsilon_{\rm B})^2\big[A_+[\epsilon_+(\psi)]
+A_+[\epsilon_-(\psi)]\big] {\rm d}\psi\Bigg\}, 
\end{eqnarray}
\end{widetext}
and we solve these equations numerically. 

\section{Estimation procedure}\label{estimation}

The secret key rate formula given by Eq.~(\ref{rate})  
can be lower bounded by
\begin{equation}\label{rate2}
R^i\geq{}qp_{\rm acc}\big\{-Q^if(E^i)H(E^i)+(p_1^iY_1+p_0^iY_0)[1-H(e_1^U)]\big\},
\end{equation}
where $e_1^U$ denotes an upper bound on the single-photon error rate $e_1$.
Hence, for our purposes, it is enough 
to obtain a lower bound on the quantities
$p^{i}_{1}Y_1+p^{i}_{0}Y_0$ for all $i\in\{s,d\}$, together with $e_1^U$. For that, we can directly use the results
obtained in Ref.~\cite{passive3}, which we include in this Appendix for completeness. 
The probabilities $p_n^i$ given by Eq.~(\ref{prob}) need to satisfy certain 
conditions that we confirm numerically. In particular, we
have that
\begin{eqnarray}\label{rome_tues1}
p^{i}_{1}Y_1+p^{i}_{0}Y_0&\geq&{}\textrm{max}\Bigg\{\frac{p^{i}_{1}(p^d_2Q^{s}-p^{s}_2Q^d)}{p^d_2p^{s}_1-p^{s}_2p^d_1}  \nonumber \\
&+&\Bigg[p^{i}_{0}
-p^{i}_{1}\frac{p^d_2p^{s}_0-p^{s}_2p^d_0} {p^d_2p^{s}_1-p^{s}_2p^d_1}\Bigg]Y_0^U,0\Bigg\},
\end{eqnarray}
where
$Y_0^U$ denotes an upper bound on the background rate $Y_0$ given by
\begin{equation}
Y_0\leq{}Y_0^U=\textrm{min}\Bigg\{\frac{E^{d}Q^{d}}{p^{d}_0e_0},\frac{E^{s}Q^{s}}{p^{s}_0e_0},1\Bigg\},
\end{equation}
with $e_0=1/2$. The single-photon error rate $e_1$ can be upper bounded as
\begin{eqnarray}\label{eq2}
e_1\leq{}e^U_1&=&\textrm{min}\Bigg\{
\frac{E^{d}Q^{d}-p^{d}_0Y_0^Le_0}
{p^{d}_1Y_1^L}, 
\frac{E^{s}Q^{s}-p^{s}_0Y_0^Le_0}{p^{s}_1Y_1^L},  \nonumber \\
&&\frac{p^{s}_0E^dQ^d
 -p^d_0E^{s}Q^{s}} 
 {(p^d_1p^{s}_0-p^{s}_1p^d_0)Y_1^L}
 \Bigg\},
\end{eqnarray} 
where 
$Y_1^L$ and $Y_0^L$ represent, respectively, a lower bound on the yield $Y_1$ and the background rate $Y_0$. These 
quantities are given by
\begin{equation}\label{eq1_thermal}
Y_1\geq{}Y^L_1=\textrm{max}\Bigg\{\frac{p^d_2Q^{s}-p^{s}_2Q^d - (p^d_2p^{s}_0-p^{s}_2p^d_0)Y_0^U} {p^d_2p^{s}_1-p^{s}_2p^d_1},0\Bigg\},
\end{equation}
and
\begin{equation}\label{q6}
Y_0\geq{}Y^L_0=\textrm{max}\Bigg\{\frac{p^d_1Q^{s}-p^{s}_1Q^d}{p^d_1p^{s}_0-p^{s}_1p^d_0},0
\Bigg\}.
\end{equation}

To evaluate these expressions we need the statistics $p_n^i$ for $n=0,1,2$. Using Eq.~(\ref{prob}), and assuming again $\theta_\Lambda=\pi/2$, 
we obtain 
\begin{eqnarray}
p^{s}_0&=&A_-(\zeta),  \\
p^{s}_1&=&\zeta[A_-(\zeta)-e^{-\zeta}(I_{1,\zeta}-L_{-1,\zeta})], \nonumber \\
p^{s}_2&=&\frac{\zeta}{2}\Big\{\zeta{}A_-(\zeta)+e^{-\zeta}\Big[\frac{2}{\pi}\Big(1-\frac{\zeta^2}{3}\Big)
+(1-2\zeta)(I_{1,\zeta} \nonumber \\
&-&L_{-1,\zeta})+\zeta(I_{2,\zeta}-L_{2,\zeta})\Big]\Big\}, \nonumber
\end{eqnarray}
and
\begin{eqnarray}
p^{d}_0&=&A_+(\zeta),  \\
p^{d}_1&=&\zeta[A_+(\zeta)-e^{-\zeta}(I_{1,\zeta}+L_{-1,\zeta})], \nonumber \\
p^{d}_2&=&\frac{\zeta}{2}\Big\{\zeta{}A_+(\zeta)+e^{-\zeta}\Big[-\frac{2}{\pi}\Big(1-\frac{\zeta^2}{3}\Big)
+(1-2\zeta)(I_{1,\zeta} \nonumber \\
&+&L_{-1,\zeta})+\zeta(I_{2,\zeta}+L_{2,\zeta})\Big]\Big\}. \nonumber
\end{eqnarray}

\section{Asymptotic passive decoy-state BB84 transmitter}\label{asymptotic}

To evaluate the secret key rate formula given by Eq.~(\ref{key}) in this scenario, 
we consider that $p_s\approx{}1$, and we assume that Alice and Bob can  
estimate the relevant parameters $Y_0$, $Y_1$, and $e_1$ perfectly. 
Moreover, we use the channel and detection models introduced in Appendix~\ref{gqber}. 
In this situation, it turns out that the yields $Y_0$ and $Y_1$ are given by
$Y_0=\epsilon_{\rm B}(2-\epsilon_{\rm B})$ and $Y_1=1-(1-Y_0)(1-\eta_{\rm sys})$. 

The single-photon error rate $e_1$ can be calculated using Eqs.~(\ref{uno})-(\ref{tres}) with
$\sigma_{i,\psi}=\ket{1_\psi}\bra{1_\psi}$. After a short calculation, we obtain
\begin{eqnarray}
e_1&=&\frac{1}{2Y_1}\Bigg\{Y_0+(1-\epsilon_{\rm B})^2\eta_{\rm sys}-\frac{4(1-\epsilon_{\rm B})\eta_{\rm sys}}
{\pi-4\Omega}\nonumber \\
&\times&\sin{\bigg(\frac{\pi}{4}-\Omega\bigg)}\Bigg\}.
\end{eqnarray}

The resulting lower bound on the secret key rate is illustrated
in Fig.~\ref{figrate} (red line) for the optimized parameters $\zeta$ and $\Omega$. 


\begin{thebibliography}{99}
\bibitem{qkd} 
N. Gisin, G. Ribordy, W. Tittel and H. Zbinden, Rev. Mod. Phys. {\bf 74}, 145 (2002); 
M. Du\v{s}ek, N. L\"utkenhaus and M. Hendrych, Progress in Optics {\bf 49}, Edt. E.
Wolf (Elsevier), 381 (2006); 
V. Scarani, H. Bechmann-Pasquinucci, N. J. 
Cerf, M. Du\v{s}ek, N. L\"utkenhaus and M. Peev, 
Rev. Mod. Phys. {\bf 81}, 1301 (2009).
\bibitem{Vernam} G. S. Vernam, J. Am. Inst. Electr. Eng. {\bf XLV}, 109 (1926).
\bibitem{bb84} C. H. Bennett and G. Brassard, Proc. IEEE Int.
Conference on Computers, Systems and Signal Processing, Bangalore,
India, IEEE Press, New York, 175 (1984).
\bibitem{decoy} W.-Y. Hwang, Phys. Rev. Lett. {\bf 91}, 057901 (2003).
\bibitem{decoy2}
H.-K. Lo, X. Ma and K. Chen, Phys. Rev. Lett. {\bf 94}, 230504
(2005).
\bibitem{decoy2b} X.-B. Wang, Phys. Rev. Lett. {\bf 94}, 230503 (2005);
X.-B. Wang, Phys. Rev. A {\bf 72}, 012322 (2005); 
X.-B. Wang, Phys. Rev. A {\bf 72}, 049908(E) (2005).
\bibitem{model2}
X. Ma, B. Qi, Y. Zhao and H.-K- Lo, Phys. Rev. A {\bf 72}, 012326 (2005).
\bibitem{decoy_e} 
Y. Zhao, B. Qi, X. Ma, H.-K. Lo and L. Qian, Phys. Rev. Lett.
{\bf 96}, 070502 (2006); 
Y. Zhao, B. Qi, X. Ma, H.-K. Lo and L.
Qian, Proc. of IEEE International Symposium on Information Theory
(ISIT'06), 2094 (2006); 
C.-Z. Peng, J. Zhang, D. Yang, W.-B. Gao,
H.-X. Ma, H. Yin, H.-P. Zeng, T. Yang, X.-B. Wang and J.-W. Pan,
Phys. Rev. Lett. {\bf 98}, 010505 (2007); 
D. Rosenberg, J. W.
Harrington, P. R. Rice, P. A. Hiskett, C. G. Peterson, R. J.
Hughes, A. E. Lita, S. W. Nam and J. E. Nordholt, Phys. Rev.
Lett. {\bf 98}, 010503 (2007); T. Schmitt-Manderbach, H. Weier,
M. F\"urst, R. Ursin, F. Tiefenbacher, T. Scheidl, J. Perdigues,
Z. Sodnik, C. Kurtsiefer, J. G. Rarity, A. Zeilinger and H.
Weinfurter, Phys. Rev. Lett. {\bf 98}, 010504 (2007); 
Z. L. Yuan,
A. W. Sharpe and A. J. Shields, Appl. Phys. Lett. {\bf 90}, 011118
(2007); 
Z.-Q. Yin, Z.-F. Han, W. Chen, F.-X. Xu, Q.-L. Wu and
G.-C. Guo, Chin. Phys. Lett {\bf 25}, 3547 (2008); J. Hasegawa, M. Hayashi, T.
Hiroshima, A. Tanaka and A. Tomita, Preprint quant-ph/0705.3081; J. F.
Dynes, Z. L. Yuan, A. W. Sharpe and A. J. Shields, Optics Express
{\bf 15}, 8465 (2007).
\bibitem{random}
J. F. Dynes, Z. L. Yuan, A. W. Sharpe and A. J. Shields, Appl. Phys. Lett. {\bf 93}, 1 (2008); 
B. Qi, Y.-M. Chi, H.-K. Lo and L. Qian, Opt. Lett. {\bf 35}, 312 (2010);
K. Hirano, T. Yamazaki, S. Morikatsu, H. Okumura, H. Aida, A. Uchida, S. Yoshimori, K. Yoshimura, 
T. Harayama and P. Davis, Opt. Express {\bf 18}, 5512 (2010);
M. F\"urst, H. Weier, S. Nauerth, D. G. Marangon, C. Kurtsiefer and H. Weinfurter,
Opt. Express {\bf 18}, 13029 (2010).
\bibitem{note} 
For simplicity, 
we will first consider the case of polarization encoding. Later in this paper, we will also examine 
phase encoding.
\bibitem{4lasers}
R. J. Hughes, J. E. Nordholt, D. Derkacs and C. G. Peterson, New J. Phys. 
{\bf 4}, 43 (2002);
C. Kurtsiefer, P. Zarda, M. Halder, H. Weinfurter, P. M. Gorman, P. R. Tapster and
J. G. Rarity, Nature {\bf 419}, 450 (2002).
\bibitem{modulator}
C. H. Bennett, F. Bessette, G. Brassard, L. Salvail and J. Smolin, J. Cryptology {\bf 5}, 3 (1992);
A. Muller, J. Br\'eguet and N. Gisin, Europhysics Lett. {\bf 23}, 383 (1993);
J. Br\'eguet, A. Muller and N. Gisin, J. Mod. Opt. {\bf 41}, 2405 (1994);
A. Muller, H. Zbinden and N. Gisin, Nature {\bf 378}, 449 (1995); 
A. Muller, H. Zbinden and N. Gisin, Europhysics Lett. {\bf 33}, 335 (1996);
P. Townsend, IEEE Photonics Tech. Lett. {\bf 10}, 1048 (1998); 
G. B. Xavier, N. Walenta, G. Vilela de Faria, G. P. Tempor\~ao, N. Gisin, H. Zbinden and J. P. 
von der Weid, New J. Phys. {\bf 11}, 045015 (2009).
\bibitem{post} N. L\"utkenhaus, Appl. Phys. B: Lasers Opt. {\bf 69}, 395 (1999);
D. Gottesman and H.-K. Lo, IEEE Trans. Inf. Theory {\bf 49}, 457 (2003);
X. Ma, C.-H. F. Fung, F. Dupuis, K. Chen, K. Tamaki and H.-K. Lo, Phys. Rev. A {\bf 74}, 032330 (2006);
V. Scarani, A. Ac\'{\i}n, G. Ribordy and N. Gisin, Phys. Rev. Lett. {\bf 92}, 057901 (2004);
B. Kraus, N. Gisin and R. Renner, Phys. Rev. Lett. {\bf 95}, 080501 (2005);
R. Renner, N. Gisin and B. Kraus, Phys. Rev. A {\bf 72}, 012332 (2005);
J. M. Renes and Graeme Smith, Phys. Rev. Lett. {\bf 98}, 020502 (2007).
\bibitem{rarity} J. G. Rarity, P. C. M. Owens and P. R. Tapster, J. Mod. Opt. {\bf 41}, 2435 (1994).
\bibitem{passive1}
W. Mauerer and C. Silberhorn, Phys. Rev. A {\bf 75}, 050305(R) 
(2007); Y. Adachi, T. Yamamoto, M. Koashi and N. Imoto, Phys. Rev. 
Lett. {\bf 99}, 180503 (2007); 
X. Ma and H.-K. Lo, New J. Phys. {\bf 10}, 073018 (2008).
\bibitem{passive2}
M. Curty, T. Moroder, X. Ma and N. L\"utkenhaus, Opt. Lett. {\bf 34}, 
3238 (2009).
\bibitem{passive3}
M. Curty, X. Ma, B. Qi and T. Moroder, Phys. Rev. A {\bf 81}, 022310 (2010). 
\bibitem{passive4}
Y. Adachi, T. Yamamoto, M. Koashi and N. Imoto, New J. Phys. {\bf 11}, 113033 (2009).
\bibitem{passiveEB} 
G. Ribordy, J. Brendel, J.-D. Gauthier, N. Gisin and H. Zbinden, Phys. Rev. A {\bf 63}, 012309 (2000);
W. Tittel, J. Brendel, H. Zbinden and N. Gisin, Phys. Rev. Lett. {\bf 84}, 4737 (2000).
\bibitem{curty} M. Curty, X. Ma, H.-K. Lo and N. L\"utkenhaus, Phys. Rev. A {\bf 82}, 052325 (2010).
\bibitem{sfg1} R. W. Boyd, {\it Nonlinear Optics} (Academic Press, Ed. Elsevier Inc., 2008);
Y. R. Shen, {\it The Principles of Nonlinear Optics} (Ed. Wiley-Interscience, 
1984). 
\bibitem{kumar} P. Kumar, Opt. Lett. {\bf 15}, 1476 (1990); J. Huang and P. Kumar, Phys. Rev. Lett. {\bf 68}, 2153 (1992).
\bibitem{sfg2}
Y.-H. Kim, S. P. Kulik and Y. Shih, Phys. Rev. Lett. {\bf 86}, 
1370 (2001); 
B. Dayan {\it et al.}, Phys. Rev. Lett. {\bf 94}, 043602 (2005);  
A. Pe'er {\it et al.}, Phys. Rev. Lett. {\bf 94}, 073601 (2005); F. Zah 
{\it et al.}, Opt. Express {\bf 16}, 16452 (2008); 
S. Tanzilli {\it et al.}, Nature (London) {\bf 437}, 116 (2005); R. T. 
Thew {\it et al.}, Appl. Phys. Lett. {\bf 93}, 071104 (2008). 
\bibitem{sfg3} 
N. Sangouard {\it et al.}, Phys. Rev. Lett. {\bf 106}, 120403 (2011).
\bibitem{note_last} Such PBS transmits $-45^\circ$ linear polarization and reflects $+45^\circ$ linear polarization \cite{kok_last}.
These two orthogonal linear polarizations have creation operators
$a^\dagger_{\pm45^\circ}=1/\sqrt{2}(a^\dagger_{H}\pm{}a^\dagger_{V})$.
\bibitem{kok_last}
P. Kok {\it et al.}, Rev. Mod. Phys. {\bf 79}, 135-174 (2007). 
\bibitem{sec_bb84b}
D. Gottesman, H.-K. Lo, N. L\"utkenhaus and J. Preskill, Quantum Inf. Comput. {\bf 4}, 
325 (2004). 
\bibitem{hkl} H.-K. Lo, Quantum Inf. Comput. {\bf 5}, 413 (2005).
\bibitem{notetor} It can be shown that 
the single-photon signals emitted by the passive source illustrated
in Fig.~\ref{figure_general}, averaged over the values of Alice's key 
bit, are basis-independent \cite{curty}.
\bibitem{eff} 
H.-K. Lo, H. F. C. Chau, and M. Ardehali, J. Cryptology {\bf 18}, 133 
(2005).
\bibitem{gb}
G. Brassard and L. Salvail, in {\it Advances in Cryptology 
EUROCRYPTÕ93, Lecture Notes in Computer Science} {\bf 765}, 
edited by T. Helleseth (Springer, Berlin, 1994).
\bibitem{linear}
M. S. Bazaraa, J. J. Jarvis, and H. D. Sherali, {\it Linear 
Programming and Network Flows} (Wiley, New York, 2004), 3rd Ed.
\bibitem{Bessel} G. Arfken, {\it Mathematical Methods for Physicists} (Academic Press, New York, 1985), 
3rd Ed; M. Abramowitz and I. A. Stegun, 
{\it Handbook of Mathematical Functions with Formulas, Graphs, and Mathematical Tables}
(Dover, New York, 1972), 9th Ed.

\end{thebibliography}
\bibliographystyle{apsrev}

\end{document}